\documentclass{aastex63}
\usepackage{lineno}
\usepackage[utf8]{inputenc}
\usepackage{graphics,graphicx}
\usepackage{amsmath,amssymb,amsfonts,latexsym,cancel}
\graphicspath{{./}{figures/}}

\begin{document}

\title{The structure and stability of massive hot white dwarfs}

\correspondingauthor{S\'ilvia P. Nunes}
\email{silviapn@ita.br}

\author{S\'ilvia P. Nunes}
\affiliation{Departamento de F\'isica, Instituto Tecnol\'ogico de Aeron\'autica, Centro T\'ecnico Aeroespacial,\\ 12228-900 S\~ao Jos\'e dos Campos, S\~ao Paulo, Brazil}

\author{Jos\'e D. V. Arba\~nil}
\affiliation{Departamento de Ciencias, Universidad Privada del Norte,\\ Avenida el Sol 461 San Juan de Lurigancho, 15434 Lima,  Peru}
\affiliation{Facultad de Ciencias F\'isicas, Universidad Nacional Mayor de San Marcos,\\ Avenida Venezuela s/n Cercado de Lima, 15081 Lima,  Peru}

\author{Manuel Malheiro}
\affiliation{Departamento de F\'isica, Instituto Tecnol\'ogico de Aeron\'autica, Centro T\'ecnico Aeroespacial,\\ 12228-900 S\~ao Jos\'e dos Campos, S\~ao Paulo, Brazil}



\begin{abstract}

We investigate the structure and stability against radial oscillations, pycnonuclear reactions, and inverse $\beta$-decay of hot white dwarfs. We regard that the fluid matter is made up for nucleons and electrons confined in a Wigner-Seitz cell surrounded by free photons. It is considered that the temperature depends on the mass density considering the presence of an isothermal core. We find that the temperature produces remarkable effects on the equilibrium and radial stability of white dwarfs. The stable equilibrium configuration results are compared with white dwarfs estimated from the Extreme Ultraviolet Explorer Survey and Sloan Digital Sky Survey. We derive masses, radii, and central temperatures for the most massive white dwarfs according to surface gravity and effective temperature reported by the survey. We note that these massive stars are in the mass region where the general relativity effects are important. These stars are near the threshold of instabilities due to radial oscillations, pycnonuclear reaction, and inverse $\beta$-decay. Regarding the radial stability of these stars as a function of the temperature, we obtain that the radial stability decreases with the increment of central temperature. We also obtain that the maximum mass point and the zero eigenfrequencies of the fundamental mode are determined at the same central energy density. Regarding low-temperature stars, the pycnonuclear reactions occur in almost similar central energy densities, and the central energy density threshold for inverse $\beta$-decay is not modified. For $T_c\geq1.0\times10^{8}\,[\rm K]$, the onset of the radial instability is attained before the pycnonuclear reaction and the inverse $\beta$-decay.

\end{abstract}



\section{Introduction} \label{sec:intro}

\subsection{Equilibrium configuration of white dwarfs}

In the theory of evolution, it is known that stars that leave the main sequence with masses below $\sim10M_\odot$ end on white dwarfs \citep{Weidemann_1983, Shapiro}. These stars start their lives at high temperatures; build them by a core -usually composed of oxygen, helium, and carbon-  surrounded by an envelope that could be rich in hydrogen \citep{Dufour_2008}.

One of the first theoretically studies about white dwarfs with temperature was developed by \cite{Marshak_1940}. Considering a star constituted by an isothermal core, composed of degenerate matter, and an envelope which temperature distribution depends on energy-generation rate, made-up by non-degenerate matter, he got an estimate of the quantity of hydrogen (in mass percentage) in the envelope of the stars Si\-rius B and Eridani B. At low densities, the change from degenerate to non-degenerate matter is a perfect environment to implement temperature distribution and energy transport \citep[see, e.g., ][]{Koester_1972}, since this is associated with the effect of the electron temperature.

Usually, in the envelope, it is considered that the energy transport mechanism is realized by radiation or by convection.  In the first case, the energy transport by radiation is produced by photons, and some models associate luminosity to small nuclear reactions in the envelope, such as p-p and CNO cycles \citep[see, e.g.,][]{Bethe_1938,Bethe_1939,Bethe_Marshak_1939}. In the second case, the energy transport by convection appears owing to the tempe\-ra\-tu\-re difference between the core and envelope. This process has been considered in some white dwarf studies, for ins\-tan\-ce, by \cite{Vanhorn_1970,Bohm_1968,Bohm_1970,Koester_1972,Fontaine_1976,Hubbard_1970}. In both cases, energy transport provides a temperature distribution that affects the equilibrium configuration of white dwarfs. 

The temperature influence on the structure of white dwarfs has been investigated under diverse conditions. Among them, for example, in the Newtonian framework, to optimize the temperature effects, \cite{Vavrukh_2012} assume the thermal energy proportional to the kinetic energy of the electrons. Considering the Fermi-Dirac equation of state (EOS)-  with the \cite{Sommerfeld_1928} expansion- authors found that the static equilibrium configurations derived from their model are within the results estimated by the observational data. 

A generalization of the work developed in \citep{Vavrukh_2012} was published by \cite{deCarvalho_2014}. Inspired in \citep{Rotondo_2011}, once generalized the Feynman-Metropolis-Teller treatment of compressed matter to the case of the finite temperature in a Fermi-Dirac EOS, de Carvalho {\it et al.} investigated the white dwarf equilibrium configurations at finite temperatures. They found that the correction in the lattice has more influence at low masses. The authors deduced that the onset of the $\beta$-inverse decay instability is no altered for temperatures $T\lesssim10^8{\rm K}$, however, larger temperatures could have significant influences in the pycnonuclear reaction rates within white dwarfs. In addition, they also reported that the presence of temperature increases the electric field on the surface of the core  of these objects. Based on the EOS derived in \citep{deCarvalho_2014}, \cite{Boshkayev_2016} analyzed the equilibrium configuration of rotating white dwarf at finite temperature. He reported that the impact of finite temperatures is relevant in low-mass white dwarfs, like so in the estimation of the radii of these objects. Comparing his results with observational data from the Sloan Digital Sky Survey (SDSS) Data Releases $4$ \citep{Tremblay_2011,Nalez_2004,Koester_2019,Madej_2004}, the author determined that his model can be used to illustrate some white dwarfs detected from the SDSS. 

\subsection{On the stability of white dwarfs}

In the study of white dwarfs, an interesting physical property to be analyzed is their stability against some perturbations. In this sense, it is important to study and analyze how this aspect changes against small radial perturbations, pycnonuclear reaction, and inverse $\beta$-decay since they could give some information about the conditions which allow the existence of white dwarfs. 

Since the analysis of the stability of compact objects against small radial perturbations developed in  Chandrasekhar's seminal works \citep{chandrasekhar_PRL, chandrasekhar_rp}, several articles investigated the radial stability of white dwarfs at zero and nonzero temperature, without taking into account the Wigner-Seitz cell. To name a few of them, at zero temperature, in \citep{Meltzer_1966} the periods and e-folding of the lowest three normal radial pulsation of white dwarf equilibrium configurations at the end point of thermonuclear evolution are investigated. The authors found that between the mass densities $2.5\times10^8<\rho<1.3\times10^9\,[\rm g\,cm^{-3}]$, there are metastable white dwarfs with e-folding time $\gtrsim10^{10}$ years, which corresponds to the Hubble time. The radial stability of zero-temperature white dwarfs was also investigated taking into account different values of the mean molecular weight per electron, $\mu_e$, in \citep{Wheeler_1968} and for two different adiabatic indexes in \citep{Chanmugam_1977}. In these works were found a limit to the radial stability for complete degenerate white dwarfs at central mass densities $\rho_c\lesssim 10^{10}[\rm g\,cm^{-3}]$. In turn, at nonzero temperature, white dwarfs were analyzed by using an analytical approximated EOS for the relativistic Fermi gas \citep{Bisnovatyi_1966}. In this work, Bisnovatyi-Kogan obtained the critical mass of isothermal white dwarf depended on the central density. Using Chandrasekhar's equation for radial oscillations, \cite{Baglin_1966} reported that the relativistic effect can affect the stability of white dwarfs with low temperatures. Moreover, the author shows that radial instability is attained in smaller energy densities than those obtained from the classical framework.

Since the high-density matter in compact objects must be in chemical equilibrium against nuclear reactions, it is also important to investigate the white dwarfs' stability against both pycnonuclear and inverse $\beta$-reaction.

Pycnonuclear reactions,  as schematically expressed by 
\begin{equation}
^A_{Z}Y+^A_{Z}Y\rightarrow ^{2A}_{2Z}K,    
\end{equation}
have been studied for white dwarfs by \cite{Gasques_2005}. The authors developed a phenomenological formalism for pycnonuclear reaction rates between identical nuclei and applied to the carbon fusion reaction. They also found a limit for carbon burning importance of $T\sim(4-15)\times10^8\,[\rm K]$ for $\rho \lesssim 3 \times 10^9 [\rm g~cm^{-3}]$, and by the mass density $\rho \sim(3-50)\times10^9[\rm g~ cm^{-3}]$ for $T\lesssim 10^8\,[\rm K]$. This type of reaction could be interpreted like a previous event to supernova Ia explosion \citep{Niemeyer_1997,Hillebrandt_2000,Han_2004,Liu_2013,Baron_2014}.

The inverse $\beta$-reaction consists of the instability reached due to the decay of atoms A$(N,Z)$ into $A(N,Z-1)$, being $N$ the mass number and $Z$ the atomic number. This type of reaction was investigated in the context of white dwarfs, e.g., by \cite{Rotondo_2011} and by \cite{Mathew_2017}. In the first article, in the context of general relativity, the authors determined that the inverse $\beta$-reaction occurs above threshold density for white dwarfs estimated. In the second work,  it is found showing that the heavier the atom element, the less the instability threshold density.

\subsection{Our aim}

In this article, in the framework of General Relativity (GR), we study the static structure configuration and stability against small radial perturbation, pycnonuclear reaction and inverse $\beta$-decay of white dwarfs with finite temperature. Inspired in \citep{Timmes_1999,deCarvalho_2014,Boshkayev_2016}, we model the EOS for hot white dwarfs taking into account the Wigner-Seitz cell composed of electrons and nucleons \citep{Salpeter_1961} surrounded by free photons. We consider that these stars are constituted by an isothermal core, made of by degenerate matter, and an envelope, made by non-degenerate matter, which temperature distribution depends on the mass density  \citep[see][]{Bohm_1968,Shapiro}. This consideration is reasonable since the energy transport occurs by conduction in the core, making the temperature almost constant, and the radiation and convection create a temperature distribution in the envelope. For this model, the static equilibrium configurations are investigate through the numerical integration of the \cite{Tolman_1939,OV_1939} (TOV) equation. { We compare our results for the hot white dwarf structure with observable white dwarfs from the Extreme Ultraviolet Explorer (EUVE) and  SDSS reported in  \citep{Vennes_1997,Madej_2004,Nalez_2004,Tremblay_2011,Koester_2019}. We estimate the mass, radius and central temperature analyzing the effective temperature and gravity for massive stars with $M/M_{\odot}\geq 1.33$ from \citep{Vennes_1997}. For some central temperatures and surface gravities $\log \left(g/g_{\odot}\right)\geq 4.4$, we obtain an equation that connects the mass with surface gravity and effective temperature.} Furthermore, we investigate the eigenfrequencies of radial oscillations using the  \cite{chandrasekhar_PRL} pulsation equation.  We correlate the behavior of radial eigenfrequency oscillations as a function of the star mass and central temperature with the existence of hot and massive white dwarfs. We also study the dependence of some physical white dwarf properties such the fluid pressure, energy density, mass, radius, and the fundamental mode eigenfrequency with the temperature.

This article is organized as follows: In section \ref{eos_section} the EOS is described. Section \ref{SS_and_RS_equations} presents both the stellar equilibrium equations and the radial stability equations and their boundary conditions. In section \ref{pyc_beta} is depicted the pycnonuclear reactions and inverse $\beta$-decay inside white dwarfs. The results are displayed in section \ref{results}. Finally, we conclude in section \ref{conclusions}. Throughout this article is considered the units $c = 1 = G$, where $c$ and $G$ represent the speed of light and the gravitational constant. 

\section{The Equation of state}\label{eos_section}

{ The first assumption about the matter that makes up white dwarfs came from Chandrasekhar’s works \citep{Chandrasekhar_1931, Chandrasekhar_1935} with the theory of the degenerate electron-gas. These pioneering works were extended including the electrostatic energy by \cite{Auluck_1959} and inserting Thomas-Fermi deviations from uniform charge distribution of the electrons and exchange energy and spin-spin interactions between the electrons \citep{Salpeter_1961}, which was used to determine white dwarfs’ mass and radius in \citep{Hamada_1961}. According to \cite{Hamada_1961}, the electron density of the stars was affected by these implementations, decreasing Chandrasekhar's mass limit. General relativity and temperature effects were investigated concerning the stability of white dwarfs by \cite{Bisnovatyi_1966}, where critical density and temperature as a function of the star mass were obtained. In this work, it was already concluded that the critical density for electron-capture reaction responsible for the neutralization of white dwarfs should appear before the threshold density of general relativity instability. 

The stability of white dwarfs with Salpeter's correction was studied by \cite{Wheeler_1968} considering the pycnonuclear reactions and the electron capture reactions, thermonuclear processes, and radial oscillations. The effects of the magnetic field and rotation in white dwarfs were investigated by \cite{Ostriker_1968} and found that moderated magnetic fields can increase the radius of white dwarfs. Another approach for the high-density matter EOS was proposed by \cite{FMT_1949} and used to investigate the radial oscillation and stability of white dwarfs by \cite{Chanmugam_1977}. \cite{Lai_1991} in the beginning '$90$s also investigated the effect of strong magnetic fields in the degenerate EOS, and in particular, in the Baym-Pethick-Sutherland EOS \citep{BPS} used in the neutron star crust. An important review that takes into account all the progress made up to the '$90$s concerning white dwarfs' structure and EOS and also physical processes in the non-degenerate envelope was written by \cite{Koester_1990}.  

A very complete review regarding the EOS of white dwarfs and neutron stars was done by  \cite{Balberg_2000} where condensed matter at extreme densities was discussed. Recently, the \cite{FMT_1949} approximation has been applied to many works in white dwarfs. To name a few, the investigation of the white dwarf matter in \citep{Ruffini_2000,Bertone_2000}, relativistic corrections \citep{Rotondo_2011}, and the inclusion of temperature in the corrections \citep{deCarvalho_2014}. Inspired by \cite{Lai_1991} work on neutron stars, the white dwarfs with magnetic fields were investigated using the electron-ion interactions with a body-centered cubic lattice correction  \citep{Otoniel_2019}, and also with the addition of face-centered cubic, simple cubic lattice, and hexagonal close-packed in \citep{Chamel_2014}. Since the EOS used by Salpeter allows us to obtain similar white dwarfs static equilibrium configurations compared to the ones obtained using the other EOSs reported in the previously mentioned works, we decided to use Salpeter’s approach. In such a way, the total pressure that supports the white dwarf against the collapse and the total energy density are respectively considered as follow:
\begin{eqnarray}
P&=&P_L+P_R+P_{i}+P_{e},\label{eqeos1}\\
\varepsilon&=&\varepsilon_{L}+\varepsilon_R+\varepsilon_{i}+\varepsilon_{e},\label{eqeos3}
\end{eqnarray}
where the subscripts $L$, $R$, $i$, and $e$ indicate respectively the pressure and density of the lattice (in \citep{Salpeter_1961} correction), radiation, nucleons, and electrons \citep{Timmes_1999}. The numerical method to obtain the  energy density and pressure contributions of Eqs.~(\ref{eqeos1}) and (\ref{eqeos3}) is explained in details in subsection \ref{num_met}.}

{ We regard that these stars are constituted of an isothermal core, made of by degenerate matter, and an envelope, made by non-degenerate matter, which temperature distribution depends on the mass density of the form \citep{Bohm_1968,Shapiro}
\begin{equation}\label{temperature_distribution}
T/\rho^{2/3}={\rm constant}.    
\end{equation}}
{ This temperature-density distribution is valid only for densities below the threshold of degeneracy. The temperature-density distribution obtained from the envelope models of carbon white dwarfs found by Kritcher and collaborators \citep{Kritcher_2020} is different from the relation in Eq.~(\ref{temperature_distribution}). We tested the temperature-density profiles of \citep{Kritcher_2020}  in our EOS and the results are very similar to the ones obtained in our work with Eq.~(\ref{temperature_distribution}). In fact,  the sensitivity in the EOS of these temperature-density profiles is quite small since in the envelope the ranges of temperatures and densities are quite small compared to the core ones.}

\section{Stellar equilibrium equations and radial perturbation equations}\label{SS_and_RS_equations}

\subsection{The energy-momentum tensor and the background line element}

The fluid makes up the white dwarfs is depicted by the perfect energy-momentum tensor, which can be represented in the form
\begin{equation}\label{tem}
T_{\mu\nu}=\left(p_0+\varepsilon_0\right)u_{\mu}u_{\nu}+p_0\,g_{\mu\nu}.
\end{equation}
$u_{\mu}$ and $g_{\mu\nu}$ stand the fluid’s four velocity and the metric tensor, respectively.

The unperturbed line element employed to investigate the equilibrium configuration of hot white dwarfs is of the form:
\begin{equation}
ds^2=-e^{\nu_0} dt^2+e^{\lambda_0} dr^2+r^2(d\theta^2+\sin^2\theta d\phi^2 ),
\end{equation}
where $(t, r, \theta, \phi)$ being the Schwarzschild-like coordinates.

The variables of the fluid $P_0$ and $\varepsilon_0$, and of the metric $\nu_0$ and $\lambda_0$ are functions of the coordinates $t$ and $r$. When radial stability against small radial perturbations is investigated, following the Chandrasekhar method, the functions aforementioned with subscript $0$ can be divided into the form \citep{chandrasekhar_PRL}:
\begin{equation}\label{descomposicao}
f_0(t,r)=f(r)+\delta f(t,r).
\end{equation}
$f(r)$ denotes the physical quantities of the fluid and the unperturbed metric functions. $\delta f(t,r)$ represents the Eurelian perturbations which depend on the coordinates $t$ and $r$.

\subsection{Stellar equilibrium equations}

The stellar equilibrium equations employed to investigate the configuration of white dwarfs in a static regime, i.e., in an unperturbed system, $\delta f(t,r)=0$, are placed as following
\begin{eqnarray}
&&\frac{dm}{dr}=4\pi\varepsilon r^2 ,\label{eq_masa}\\
&&\frac{dp}{dr}=-(p+\varepsilon)\left(4\pi rp+\frac{m}{r^2}\right)e^{\lambda},\label{tov2}\\
&&\frac{d\nu}{dr}=-\frac{2}{(p+\varepsilon)}\frac{dp}{dr},\label{eq_nu}
\end{eqnarray}
with
\begin{equation}\label{eq_lambda}
e^{\lambda}=\left(1-\frac{2m}{r}\right)^{-1}.
\end{equation}
As is habitual, the function $m$ represents the mass within a sphere of a radius $r$. Equation \eqref{tov2} is known as the hydrostatic equilibrium equation, also called as TOV equation.

Once known the EOS, Eqs.~\eqref{eqeos1} and \eqref{eqeos3}, with the aim to look for equilibrium solutions, we solve simultaneously the stellar equilibrium equations (Eqs.~\eqref{eq_masa}-\eqref{eq_nu}).  This set of equations are integrated from the center ($r=0$) to the star's surface ($r=R$). The initial conditions at the center of the star are
\begin{equation}\label{TOV_IC}
m(0)=0,\quad \varepsilon(0)=\varepsilon_c,\quad T(0)=T_c,\quad{\rm and}\quad \nu(0)=\nu_c.
\end{equation}
The star's surface is found when
\begin{equation}\label{pressure_surface}
P(R)=0,
\end{equation}
and consequently $T(r=R)=0$. At this point, the interior solution connects smoothly to the spacetime outside the star. This indicates that the interior and exterior metric functions are related as follow:
\begin{equation}\label{TOV_BC}
e^{\nu(R)}=e^{-\lambda(R)}=1-\frac{2M}{R},
\end{equation}
where $M$ represents the total mass of the star. In addition, this relation provides the boundary condition for the functions $\nu$ and $\lambda$ at the star's surface.

\subsection{Radial perturbation equations}

To investigate the stability of white dwarf against small radial perturbations, it is necessary to determine the radial oscillation equations. For such an aim, firstly, the Eulerian perturbations must be derived. Once the nonzero four-velocity components are defined, the fluid and spacetime variables are decomposed into the form presented in Eq.~\eqref{descomposicao}. After replacing these definitions and decompositions in the field equations, the Eulerian perturbations are found just keeping the first-order terms. 

The radial pulsation equation is derived taking into account the linearized form of the stress-energy tensor conservation, the Eulerian perturbations, and considering that the perturbed quantities have a time dependence $e^{i\omega t}$; where $\omega$ is called as eigenfrequency. This constant parameter is determined through the equality \citep{chandrasekhar_PRL,Baglin_1966}
\begin{equation}\label{eq17}
    \omega^2=\frac{\mathcal{Z}}{\mathcal{D}}.
\end{equation}
The functions $\mathcal{Z}$ and $\mathcal{D}$ are respectively determined by the equations
\begin{eqnarray}
&&\mathcal{Z}=4\int^R_0 e^{\nu+\lambda/2}r^3\frac{dp}{dr}dr+8\pi\int^R_0 e^{\nu+3\lambda/2}p(p+\varepsilon)r^4dr+9\int^R_0 e^{\nu+\lambda/2}r^2\Gamma pdr-\int^R_0 \frac{e^{\nu+\lambda/2} r^4}{(p+\varepsilon)}\left[\frac{dp}{dr}\right]^2dr,\label{eq18}
\\
&&\mathcal{D}=\int^R_0 {e^{3\lambda/2}}(p+\varepsilon){r^4}dr,\label{eq19}
\end{eqnarray}
where $\Gamma$ represents the adiabatic index and it is determined by the condition:
\begin{equation}
\Gamma=\frac{d\log p}{d\log\varepsilon}.
\end{equation} 

Equation \eqref{eq17}  helps us to discriminate stable equilibrium solutions from unstable ones. This is possible by analyzing the eigenfrequency values $\omega$.

\section{Pycnonuclear reactions and inverse $\beta$-decay}\label{pyc_beta}

\subsection{Pycnonuclear reactions}

Pycnonuclear reactions possibility in the dense stellar matter \citep{Yakovlev_2005} creates a constraint in the white dwarf's chemical composition. These reactions are almost independent of temperature and appear even at zero temperature. A rigorous approach to calculate these nuclear reactions was realized by  \cite{Salpeter_1969}, who established a temperature-dependent pycnonuclear rate to the zero-temperature rate. For a pure carbon core \citep{Otoniel_2019}, where $^{12}{\rm C}+^{12}{\rm C}\rightarrow\,^{24}{\rm Mg}$, this reaction rate can be written as \citep{Salpeter_1969}:
\begin{equation}\label{pycno}
\frac{R_{\rm pyc}(T)}{R_{\rm pyc}(0)}-1=a_1\lambda^{-1/2}{\left[1+a_2e^{b_1\beta^{3/2}}\right]^{-1/2}}\times\exp\left\{b_2\beta^{3/2}+\lambda^{-1/2}a_3e^{{b_1\beta^{3/2}}}\left[1-a_4e^{b_1\beta^{3/2}}\right]\right\},
\end{equation}
where $R_{\rm pyc}(0)$ represents the pycnonuclear reaction at zero temperature, and $a_1$, $a_2$, $a_3$, $a_4$, $b_1$, and $b_2$ are model-dependent dimensionless constants. 

The time to complete the atomic nuclei fusion is obtained from \citep{Gasques_2005, Boshkayev_2013,Otoniel_2019}
\begin{equation}
\tau_{\rm pyc}=\frac{n_N}{R_{\rm pyc}}.
\end{equation}
As considered in \citep{Otoniel_2019}, we employ the pycnonuclear reaction time $\tau_{\rm pyc}=10\,[\rm Gyr]$, which corresponds to an upper limit where pycnonuclear reactions are extremely slow \citep{Chugunov_2007}. For white dwarfs, the limit of $\rho_{\rm pyc}(10\,[\rm Gyr])$ corresponds to a maximum mass density for stable stars against pycnonuclear reactions. Through these times, it can be estimated the mass density where these reactions appear, $\rho_{pyc}(\tau_{\rm pyc})$. For white dwarfs, the limit of $\rho_{pyc}(10\,[\rm Gyr])$ corresponds to a maximum density for stable stars, since from this point pycnonuclear reactions begin to appear.

\subsection{Inverse $\beta$-decay}

It is known that the matter inside white dwarfs may experience instability against the inverse $\beta$-decay process,
\begin{equation}
A(N,Z)+e^{-}\rightarrow A(N+1,Z-1)+\nu_{e}.
\end{equation}
Due to this process, atomic nuclei turn into more neutron-rich and, as a consequence, the electron energy density and pressure are reduced thus leading to a softer EOS \citep{Gamow_1939,Shapiro}. Since we are considering a nucleus of $^{12}$C, the instabilities are reached at energies larger than $\epsilon_Z^\beta=13.370\,[\rm MeV]$ \citep[see][]{Rotondo_2011,deCarvalho_2014,Otoniel_2019}.

\section{Equilibrium and stability of hot white dwarfs}\label{results}

\subsection{Numerical method}\label{num_met}

{ Due to partial degeneracy considered in this model, the pressure and energy of electrons appearing on the EOS (Eqs.~(\ref{eqeos1}) and (\ref{eqeos3})) are solved numerically by means of the adaptive quadrature method. Through this numerical method, we reproduce the results reported by \cite{Vavrukh_2012}, which resolve the same EOS through the Sommerfeld approximation, and the results found in  \citep{deCarvalho_2014,Boshkayev_2016}, where the authors investigate the equilibrium of white dwarf with a constant finite temperature. 

Once defined the EOS, both the stellar equilibrium equations, Eqs.~\eqref{eq_masa}-\eqref{eq_nu}, and radial stability equations, Eqs. (\ref{eq18}) and (\ref{eq19}),  are integrated from the center ($r=0$) towards the surface of the spherical object ($r=R$) through the Runge-Kutta fourth-order method for different values of $\varepsilon_c$, $T_c$, and a trial value for $\nu_c$. 

The numerical solution of the stellar structure equations begins with the initial conditions \eqref{TOV_IC} at $r=0$. Once given $\varepsilon_c$, $T_c$, and $\nu_c$, the integration proceeds from the center toward the star's surface where  $P(R)=0$. Nonetheless, whether after the integration the condition Eq.~\eqref{TOV_BC} is not satisfied, $\nu_c$ is corrected through a Newton-Raphson iteration scheme until it fulfills this condition. Thus, the zero fluid pressure determines the total radius $r=R$ and total mass $M=m(R)$.

After determining the coefficients $p$, $\varepsilon$, $m$, $\lambda$, and $\nu$ for each $\varepsilon_c$, $T_c$, and correct $\nu_c$, the radial stability equations are integrated from the center to the surface of the star. After the integration, the eigenfrequency squared is found by Eq.~\eqref{eq17}.}

\subsection{Influence of temperature on the fluid pressure, energy density, and mass of the star}

\begin{figure}[h!] 
\begin{center}
\includegraphics[width=0.6\linewidth]{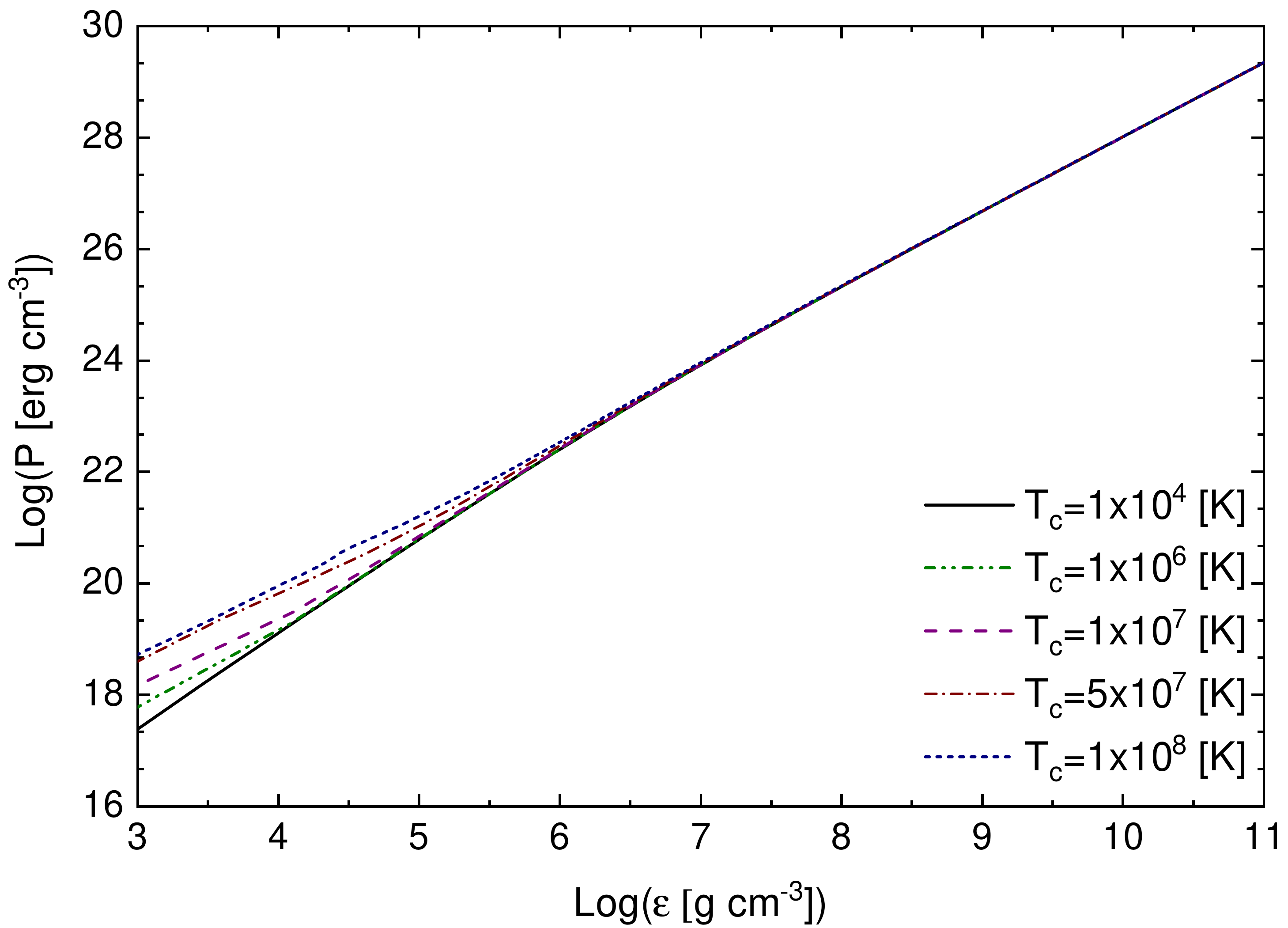}
\caption{Fluid pressure against the energy density for some different central temperatures.}
\label{figeos}
\end{center}
\end{figure}

With the purpose to observe the EOS behavior, in Fig.~\ref{figeos} is plotted the change of the fluid pressure against the energy density for some different central temperatures. The energy density employed goes to $10^3\,[\rm g ~cm^{-3}]$ to $10^{11}\,[\rm g~ cm^{-3}]$.

In Fig.~\ref{figeos}, in all cases presented, it can be observed that the pressure decays monotonically with the energy density. Moreover, the effects of central temperature are also noted in the graphic. At low energy densities, the pressure has a slight growth with central temperature. This increment is associated with the increase of the radiation pressure, and the pressure of the nucleons.

{ At the central energy density $10^4\,[\rm g\,cm^{-3}]$ and central temperatures interval $\left[10^4,10^7\right]\;[\rm K]$, we find similar fluid pressure than those ones reported by \cite{deCarvalho_2014}; namely, we derive $p_c$ in the range $[1\times10^{19},2\times10^{19}]\,[\rm dyn/cm^{2}]$.  This result can be understood since the temperature distribution is constant in the range of $\varepsilon$ and $T_c$ aforenamed, and the treatment assumed in this work (Salpeter correction) and the one employed by Carvalho (Feynman-Metropolis-Teller treatment) are similar. In addition, the central pressures found are lower than the one derived by \cite{Boshkayev_2016}. This could be associated with the fact that we are considering the Wigner-Seitz lattice correction.}


\begin{figure}[t]
\centering
\includegraphics[width=0.6\linewidth,angle=0]{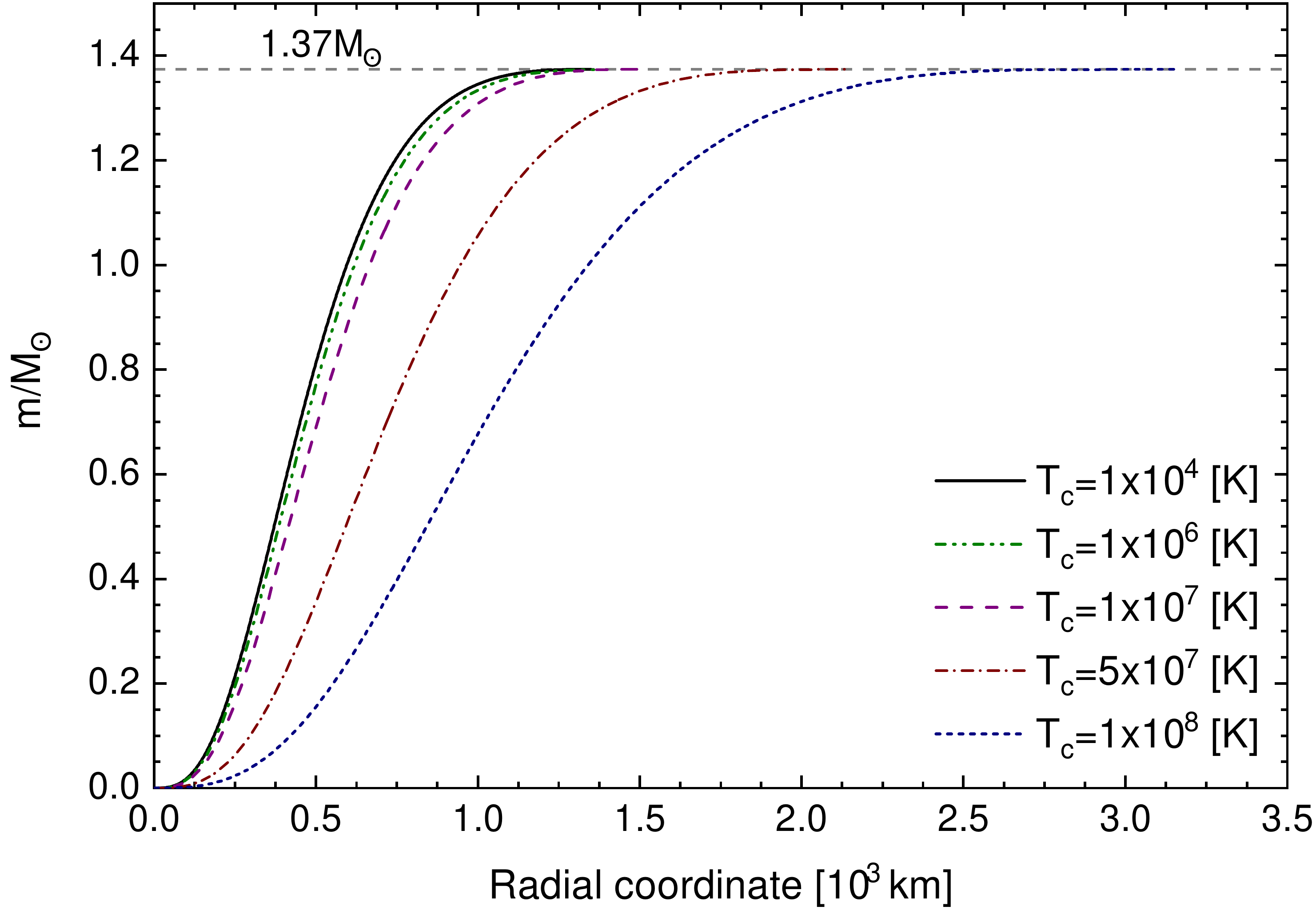}
\caption{Mass, normalized in solar masses $M_{\odot}$, as a function of the radial coordinate for five central temperature values and a total mass $M=1.37M_\odot$.}\label{top_bottom}
\end{figure}

{ The star mass as a function of the radial coordinate is presented in Fig.~\ref{top_bottom} for a white dwarf with a total mass of $1.37M_\odot$ and several central temperatures. From this figure, we conclude that for a fixed  star mass, the effect of the temperature is still important for very massive WD:  when the temperature increases the star radius increases, and this effect is more pronounced for central temperatures $T_c\geq 10^8\rm [K]$. For the extreme case of a central temperature $T_c=10^8\rm [K]$ the radius increases $135\%$ compared to $T_c=10^4\rm[K]$ due to nucleons' pressure. This result was not observed for near-Chandrasekhar mass white dwarfs with finite temperature in previous study \citep{Boshkayev_2016}, since this thermal pressure was not taken into account.  This is an important observation since very massive white dwarfs, with larger radii than the ones derived for low temperatures ($T_c\leq10^4 [\rm K]$), are an indication of lower surface gravity comparing to the ones observed for cold white dwarfs (since $g\sim 1/r$), and high central temperature. In summary, it is important to consider mass-radius relation obtained at finite temperature in the case of very massive white dwarfs, in order to obtain the correct mass and star radius from the observed surface gravity and effective temperature values.}

\subsection{Equilibrium configurations of hot white dwarfs}

\subsubsection{Hot white dwarf equilibrium configurations sequence}

\begin{figure}[h!]
\begin{center}
\includegraphics[width=0.6\linewidth]{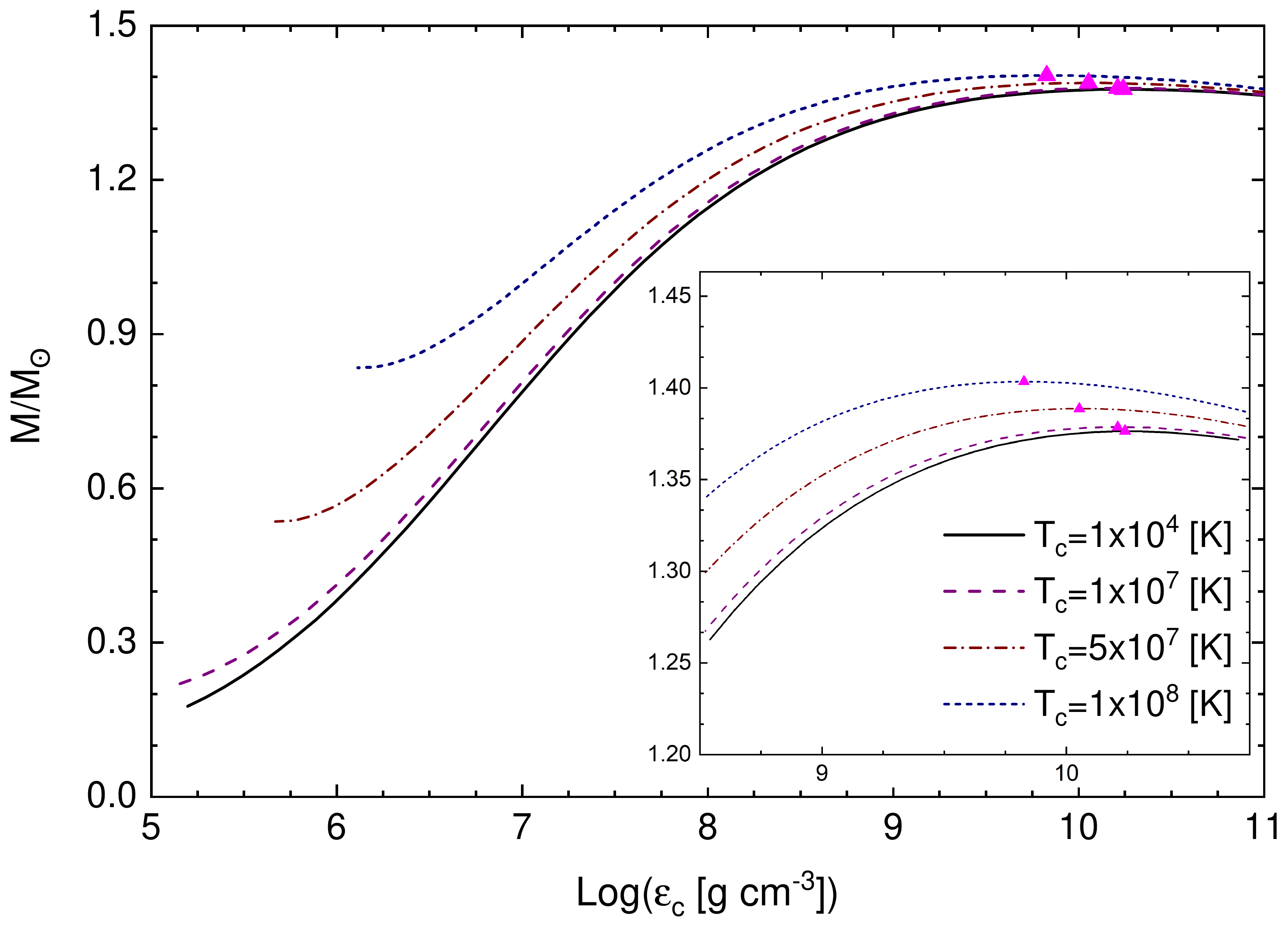}
\caption{Mass of the star, normalized to the Sun’s mass $M_{\odot}$, as a function of central energy density for four different central temperature values. The full triangles over the curves represent maximum mass points. The box within the figure shows the region where the maximum mass values are found.}
\label{fig5}
\end{center}
\end{figure}

The gravitational mass, in solar masses $M_{\odot}$, as a function of the central energy density is plotted in Fig.~\ref{fig5} for some different central temperatures. It is considered the central energy densities within the interval $10^{5}\,[\rm g\; cm^{-3}]\leq\varepsilon_c\leq10^{11}\,[\rm g\; cm^{-3}]$. The full triangles placed over the curves mark the points where the maximum masses are attained.

In Fig.~\ref{fig5}, in all central temperature considered, we note that the curves present two branches. In the first one, the mass grows monotonically with $\varepsilon_c$ until to reach the maximum mass values ($M_{\rm max}/M_{\odot}$), after this point, the curve turn-clockwise for the mass starts to decay with the growth of $\varepsilon_c$. The central energy density used to determine the maximum mass points coincide with the $\varepsilon_c$ employed to find the eigenfrequency of the fundamental mode $\omega=0$. This indicates that the maximum mass point divides the regions constituted by stable equilibrium configurations from regions established by the unstable ones. Thus, the regions make up by  stable and unstable equilibrium configurations against small radial perturbation are differentiated through the inequalities $dM/d\varepsilon_c>0$ and $dM/d\varepsilon_c<0$, respectively. These conditions are necessary and sufficient to recognize a stable region from the unstable one. 

Equilibrium configurations with energy densities lower than those considered in curves with $T_c\gtrsim5\times10^{7}[\rm K]$ were also analyzed. As well as in references \citep{deCarvalho_2014,Boshkayev_2016}, it is found that the total mass decreases with the increment of the central energy density. Moreover, these equilibrium configurations have an eigenfrequency of the fundamental mode close to zero ($\omega\sim0$).

\begin{figure}[h!]
\begin{center}
\includegraphics[width=0.6\linewidth]{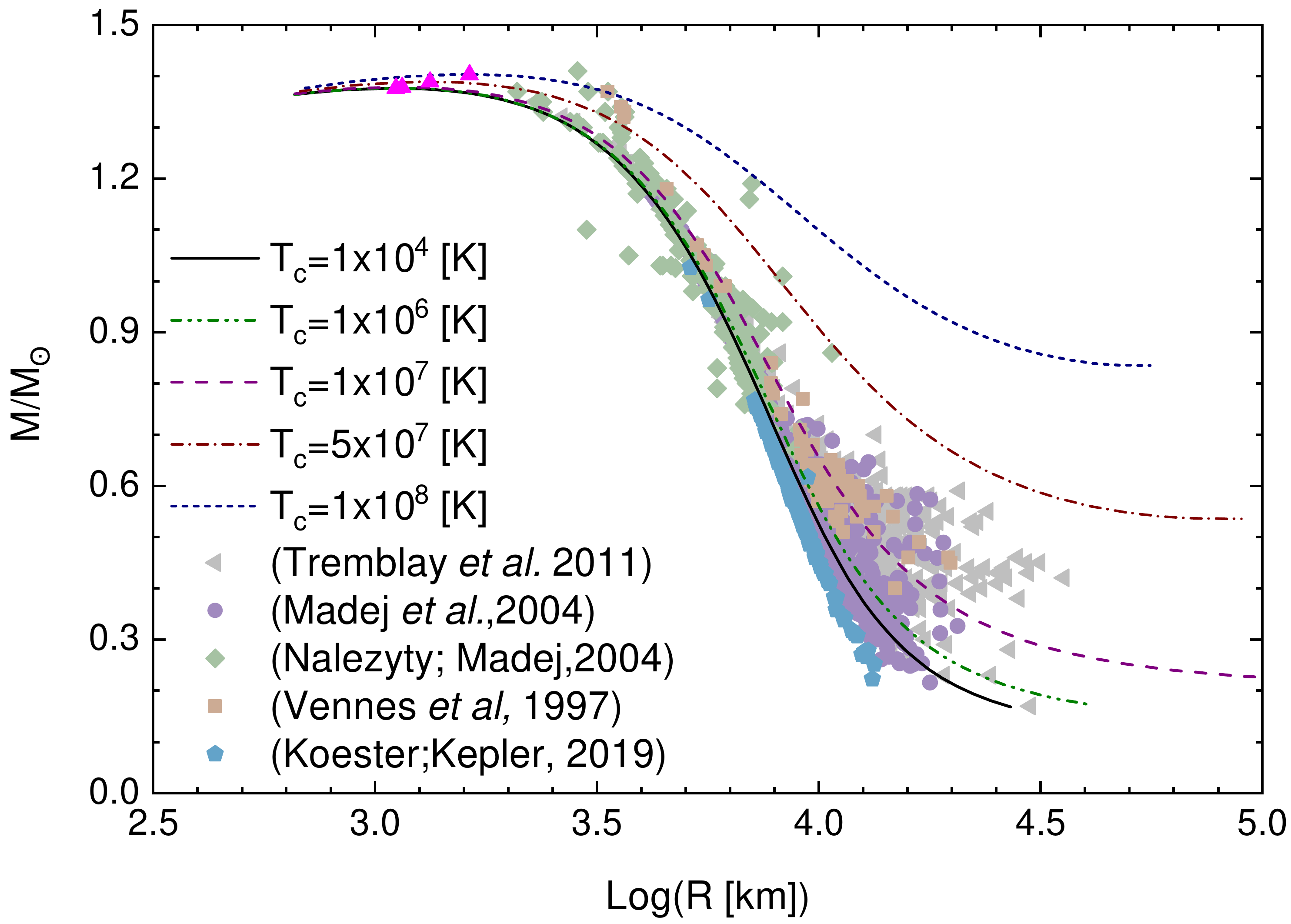}
\caption{Mass-radius curves for different central temperatures. The full triangles represent the maximum mass points. Observational data extracted from catalogs \citep{Tremblay_2011}, \citep{Nalez_2004}, \citep{Koester_2019}, \citep{Vennes_1997} and \citep{Madej_2004} are respectively marked in gray triangles, purple circles, green diamond, orange squares and blue hexagons.}
\label{fig2}
\end{center}
\end{figure}

The behavior of the total mass as a function of the radius is plotted in Figure \ref{fig2} for a few different central temperatures. The full triangles  in pink over the curves indicate the maximum mass points. The largest total radii shown in each curve are derived from the respective minimum central energy density value considered in each curve of Fig.~\ref{fig5}. As aforesaid, for lower central energy densities than those mentioned, the eigenfrequency of the fundamental mode is close to zero. Moreover, in figure, some observational results obtained from the catalogs \citep{Tremblay_2011}, \citep{Nalez_2004}, \citep{Koester_2019}, \citep{Vennes_1997} and \citep{Madej_2004} are respectively presented in gray triangles, purple circles, green diamond, orange squares, and blue hexagons.

The central temperature influence on the total mass and radius is noted in Fig.~\ref{fig2}. In all curves, the mass grows monotonically with the diminution of the total radius until it reaches $M_{\rm max}/M_\odot$. After this point, the $M(R)$ curves turn anticlockwise to the masses start to decrease with the radii diminution. In figure, a large group of the white dwarfs detected is placed below $M=1.3\,M_{\odot}$ and $T_c=10^7\,[\rm K]$ and a small group is located in higher masses and central temperatures. These results could indicate that the mass of a white dwarf is associated with the central temperature in their cores. I.e.,  for higher central temperatures, more massive white dwarfs are found. The growth of the mass with the temperature could be understood noting that some factors that compose the total fluid pressure - the pressures coming from by the radiation, and nucleons - increase with $T_c$. This increment in the pressure helps to support more mass against the collapse. { Near the maximum-mass limit, we have noticed that some white dwarfs observed are within the range of mass with high central temperature. In particular, some of them present a small gravity value compared to the obtained for low-temperature white dwarfs.}

\begin{figure}[h!] 
\begin{center}
\includegraphics[width=0.6\linewidth]{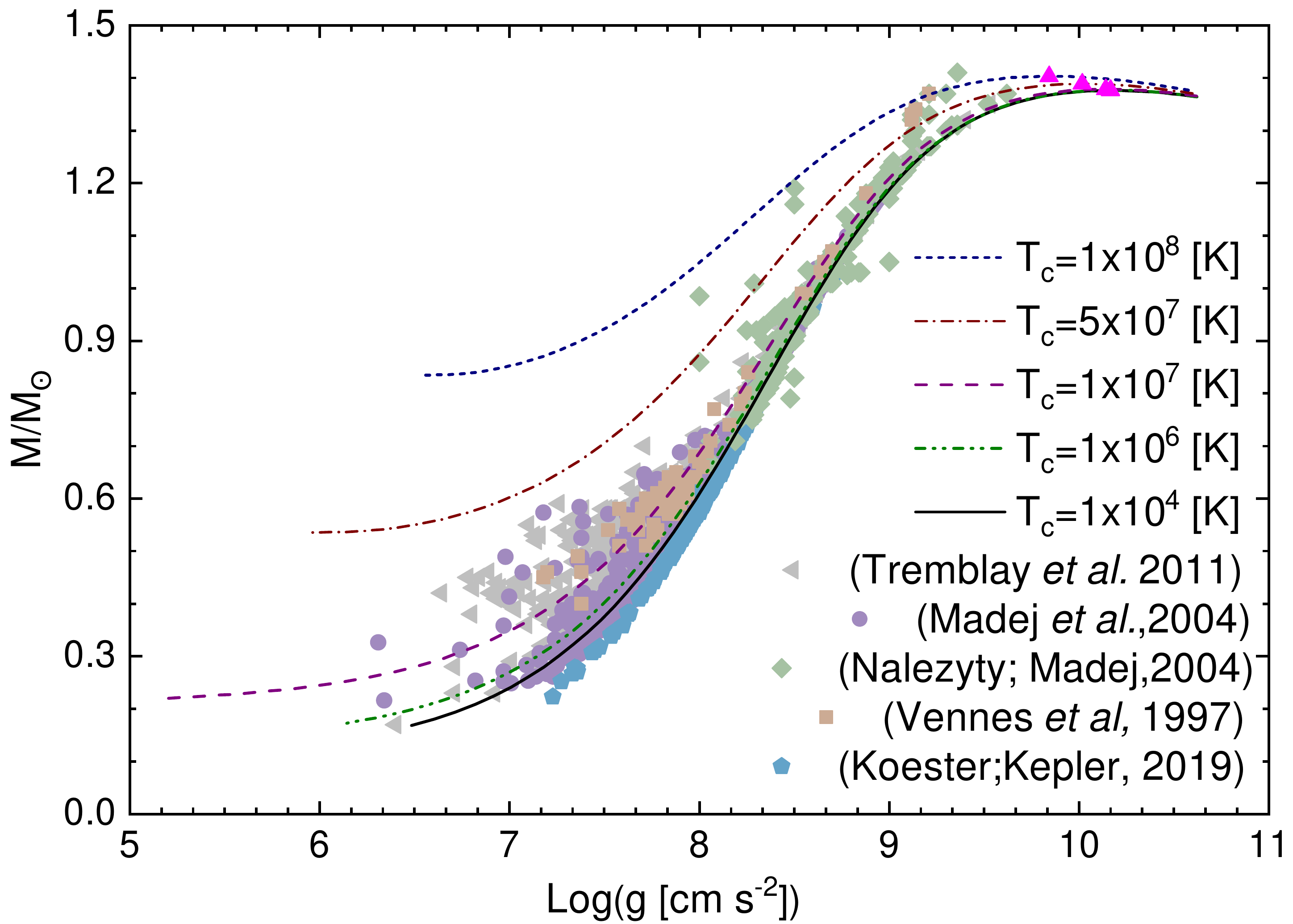}
\caption{The surface gravity of the white dwarf against its total gravitational mass, for some central temperatures. The full triangles over the curves mark the maximum mass points. Observational results obtained from catalogs \citep{Tremblay_2011}, \citep{Nalez_2004}, \citep{Koester_2019}, \citep{Vennes_1997} and \citep{Madej_2004} are respectively indicated in gray triangles, purple circles, green diamonds, orange squares and blue hexagons.}
\label{fig1}
\end{center}
\end{figure}

{ Comparing our results with the ones reported by \cite{Boshkayev_2016}, for a star made up by $^{12}\rm C$ and $\mu_e=2$, with a central energy density $10^{6} \rm ~[g\,cm^{-3}]$ and temperature $T=10^6\,[\rm K]$, it is found that the mass of the star is smaller by $7\%$ and the radius large by $23\%$ and, for a temperature of $T=10^7\,[\rm K]$, it is derived nearly the same total mass and a radius greater by $34\%$. As can be seen, the mass derived in this work is in good agreement with the one reported by \cite{Boshkayev_2016}, unlike the total radius which differs significantly. The radii derived are larger due to the inclusion of the nucleon's pressure, thermal energy, lattice corrections, and electrons energy contributions in the EOS. In the maximum total masses range, we find that the white dwarf's masses depend on central temperature; unlike the ones published in \citep{Boshkayev_2016}, where the total masses stay independent of the temperature. For instance, for $T_c=10^8 ~\rm[K]$, we find the maximum mass value $1.40 M_\odot$ and its respective total radius $1400\,[\rm km]$, while Boshkayev found white dwarf with a mass of $1.43M_\odot$ and radius $1000\,[\rm km]$. This difference is associated with the nucleons' pressure effects, which are still important for massive white dwarfs, and not only for the low mass ones. It is the first time that the nucleons’ pressure is taken into account in the study of massive white dwarfs; it was only considered in the analysis of low-mass white dwarfs  \citep{deCarvalho_2014}.}

Gravity at the white dwarf surface versus its gravitational mass is shown in Figure \ref{fig1} for different central temperatures. The full triangles over the curves mark the maximum mass values. Some observational data  taken from catalogs \citep{Tremblay_2011}, \citep{Nalez_2004}, \citep{Koester_2019}, \citep{Vennes_1997} and \citep{Madej_2004} are respectively marked in gray triangles, purple circles, green diamond, orange squares and blue hexagons. In all cases, the total mass grows monotonically with the gravity until it attains a maximum mass value; after this point, the curve turn-clockwise so that the mass starts to decrease with the increment of gravity.

\begin{deluxetable*}{cccc}
\tablenum{1}
\tablecaption{The central temperature employed with the maximum masses with their respective total radii and central energy density. \label{table3data}}
\tablewidth{0pt}
\tablehead{
\colhead{$T_c$} & \colhead{$M_{\rm max}$} & \colhead{$R$} & \colhead{$\varepsilon_c$}  \\
\colhead{[K]} & \colhead{$M_\odot$} & \colhead{[km]} & \colhead{$[\rm g~cm^{-3}]$} 
}
\startdata
$1.0\times10^6$ & $1.3766$  & $1114.9$  & $1.7440\times10^{10}$ \\ \hline
$1.0\times10^7$ & $1.3787$  & $1150.1$  & $ 1.6216\times10^{10}$ \\ \hline
$5.0\times10^7$ & $1.3887
$  & $1334.6$  & $1.1181\times10^{10}$ \\ \hline
$1.0\times10^8$ & $1.4034$  & $1630.8$ & $6.7469\times10^{9}$ 
\enddata
\end{deluxetable*}

From Figure \ref{fig1}, we can note that for a fixed value of surface gravity, the total mass increases with the central temperature. The growth of the mass with the central temperature can be explained by noting that the central pressure increases with $T_c$, from this, we understand that the central temperature acts as an effective pressure that helps the fluid pressure to support more mass against the collapse.

{ The results reported in Figs.~\ref{fig2} and \ref{fig1} are important in the cooling study of hot $T_c>10^7[\rm K]$ and very massive white dwarfs $M> 1.37M_\odot$. In the common cooling process, these stars shrink, thus maintaining constant the baryonic mass \citep{Althaus_2010} and increasing their densities. In our curves with constant gravitational mass, this would occasion them to run from a stable region to an unstable one (according to the threshold of instabilities due to radial oscillations, pycnonuclear reaction, and inverse $\beta$-decay, see Section \ref{section_stability}). This evolutionary instability could explain the star collapse and justify an interesting mass limit for observable white dwarfs. Besides, as we consider the gravitational mass and not the baryonic one for this analysis, such collapse would not occur if they lose enough gravitational mass, thus allowing a stable cooling. To obtain a robust conclusion regarding this possible mechanism originating type Ia supernovae explosion further investigations are needed.}

In Table \ref{table3data}, the central temperatures $T_c$ used and the maximum white dwarf masses values and their respective radii and central energy densities are shown. It is found that for central temperatures range {$10^4 < T_c\lesssim 10^{7}[\rm K]$}, near the maximum total mass, white dwarf masses remains nearly constant. At this range, the total pressure maintain nearly constant with the increment of temperature; since the electron pressure decays with $T$, and the contribution of the radiation pressure and pressure of nucleons do not contribute considerably to the white dwarf's structure. In turn, for $T_c\gtrsim10^{7}~[\rm K]$, an increase in total mass is observed. At this central temperature range, the contribution of $P_R$ and $P_N$ produce considerable effects at the white dwarf's structure. Thus, the white dwarf's mass at $T_c\gtrsim10^{7}~[\rm K]$ is larger than $M/M_{\odot}$ at {$10^4 < T_c\lesssim 10^{7}[\rm K]$}. On the other hand, in all cases analyzed here, at the maximum masses, it is found that their respective total radii and central energy densities increase and decrease with temperature, respectively.

\subsubsection{The procedure to identify central temperatures in massive white dwarfs}\label{massive}

{ The existence of some very massive white dwarfs have been shown by \cite{Vennes_1997}. According to their analysis, they attained the effective temperature, and gravity and used a fitting to estimate the mass of such stars. In order to verify our model, we use the effective temperature and gravity reported in \citep{Vennes_1997} to fit our curves. In our model, the effective temperature is obtained using Stefan-Boltzmann law in the photons' luminosity, which depends on the elements' mass contribution \citep[see][]{Shapiro}.}

{ On the left and right-hand side of Table \ref{tabledata1_v2}, we report, respectively, the data from \citep{Vennes_1997} for some observational white dwarfs with $M_{He}=10^{-4}M_\odot$ and the mass, radius, central temperature, and $\alpha$; with $\alpha$ being a dimensionless parameter that relates the effective temperature, gravity, and central temperature by means of the following relation \citep{Koester_1976}}
\begin{equation}\label{eq42}
\frac{T_{eff}^4}{g}=2.05\times10^{-10}T_c^{\alpha}.
\end{equation}
{ The index $\alpha$ takes the value of $2.56$ in \citep{Koester_1976}.}

\begin{deluxetable*}{lcccc|cccccc}
\tablenum{2}
\tablecaption{Left-hand side: Massive white dwarfs observed, with their respective effective temperatures and gravity on their surfaces. Right-hand side: Mass, radius, central temperature and $\alpha$ parameter for $M_{He}=10^{-4}M_\odot$ derived by using our numerical results. \label{tabledata1_v2}}
\tablewidth{0pt}
\tablehead{
\colhead{Simbad} & \colhead{$T_{eff}$} & \colhead{$\log (g)$}& \colhead{$M $}&\colhead{$R $} \vline&   \colhead{$M $}& \colhead{$R $} &\colhead{$T_{c} $}& \colhead{$\alpha$}\\
\colhead{ Name} &\colhead{$10^4\,\rm[K]$} & \colhead{$\rm [cm/s^2]$} &\colhead{$M_\odot$} &\colhead{$10 ^3\;\rm[km]$}\vline&     \colhead{$M_\odot$} & \colhead{$10^3\;\rm[km]$}& \colhead{$10^7 \rm[K]$} &\colhead{}
}
\startdata
EUVE J$0003+43.6$     & $4.24$  & $9.30\pm0.12$ & $1.37\pm0.12$ &$3.02\pm 0.21$&$1.33\pm0.02$ & $2.97\pm0.39$ & $4.54\pm0.36$ &$2.47\pm0.03$\\ 
\hline
WD $0136+251$& $3.94$  & $9.12\pm0.13$ &   $1.28\pm0.07$  &$3.59\pm0.26$& $1.30\pm 0.03$ &$3.61\pm0.50$& $4.70\pm0.40$&$2.47\pm 0.03$                    
\\ \hline
WD $0346-011$  & $4.32$  & $9.21\pm0.05$ &$1.33\pm 0.03$ & $3.30\pm 0.09$&  $1.32\pm0.01$   &$3.28\pm 0.18$&  $4.91\pm0.16$&$2.47\pm0.01$ 
\\ \hline
EQ J$0443-037$  & $6.51$  & $9.12\pm0.12$ &$1.29\pm0.06$ & $3.60\pm 0.24$& $1.33\pm0.02$   &$3.65\pm0.48$ & $8.36\pm0.67$&$2.50\pm0.03$ 
\\ \hline
EQ J$0916-197$  & $5.64$  & $9.12\pm0.20$ &  $1.29\pm0.11$ &$3.60\pm 0.40$ &$1.32\pm0.03$   &$3.64\pm0.80$  &$7.12\pm0.94$&$2.49\pm 0.05$ 
\\ \hline
WD $1659+440$  & $3.05$  & $9.38\pm0.07$ & 
$1.41\pm0.04$ & $2.86\pm 0.11$ & $1.33\pm0.01$   &$2.70\pm0.21$&  $2.95\pm0.14$&$2.44\pm0.02$ 
\\ \hline
EUVE J$1535-77.4$  & $5.80$  & $9.14\pm0.07$ &  $1.30\pm 0.11$ & $3.54\pm 0.39$ & $1.32\pm0.03$   &$3.55\pm0.79$&  $7.26\pm0.95$& $2.49\pm0.05$
\\ \hline
\enddata
\tablecomments{The mass, radius, and central temperature are obtained by fitting the effective temperature and gravity for $M_{He}=10^{-4}M_\odot$.}
{{\bf Reference.} \cite{Vennes_1997}.}
\end{deluxetable*}

 { From the results in Table \ref{tabledata1_v2}, we can note that our mass presents values in the same range to the ones reported in \citep{Vennes_1997}, except for WD $1659+440$ since \cite{Vennes_1997} found a mass of $1.41\pm 0.04M_\odot$ and we obtain $1.33\pm 0.01 M_\odot$. This is due to mainly the general relativity effects, which impacts are more notorious at large total masses. For all stars analyzed, we note that our radius results are within the range as the ones reported in \citep{Vennes_1997}. Besides that, we find $\alpha=2.48\pm0.03$. This result is within the range of the values reported in \citep{Koester_1972}, $\alpha=2.50$, and \citep{Koester_1976}, $\alpha=2.56$. The central temperature related to these massive white dwarfs suggests that some of them like EQ J$0443-037$, EQ J$0916-197$, and EUVE J$1535-77.4$ can have a high central temperature.}

\begin{deluxetable*}{lcccc|cccccc}
\tablenum{3}
\tablecaption{Left-hand side: Massive white dwarfs observed, with their respective effective temperatures and gravity on their surfaces. Right-hand side: Mass, radius, central temperature and $\alpha$ parameter for $M_{H}=10^{-4}M_\odot$ and $M_{He}=10^{-2}M_\odot$ derived by using our numerical results.\label{tabledata1_v3}}
\tablewidth{0pt}
\tablehead{
\colhead{Simbad} & \colhead{$T_{eff}$} & \colhead{$\log (g)$}& \colhead{$M $}&\colhead{$R $} \vline&   \colhead{$M $}& \colhead{$R $} &\colhead{$T_{c} $}& \colhead{$\alpha$}\\
\colhead{ Name} &\colhead{$10^4\,\rm[K]$} & \colhead{$\rm [cm/s^2]$} &\colhead{$M_\odot$} &\colhead{$10 ^3\;\rm[km]$}\vline&     \colhead{$M_\odot$} & \colhead{$10^3\;\rm[km]$}& \colhead{$10^7 \rm[K]$} &\colhead{}
}
\startdata
EUVE J$0003+43.6$     & $4.24$  & $9.30\pm0.12$ & -&-& $1.32\pm0.02$ & $2.96\pm0.38$ & $3.00\pm0.19$ &$2.53\pm0.02$\\ 
\hline
WD $0136+251$& $3.94$  & $9.12\pm0.13$ & $1.32\pm0.08$ & $3.64\pm0.26$ & $1.28\pm0.01$  &$3.58\pm0.49$& $3.00\pm0.17$&$2.53\pm 0.02$                    
\\ \hline
WD $0346-011$  & $4.32$  & $9.21\pm0.05$ & $1.37\pm 0.03$ &$3.34\pm 0.09$ & $1.30\pm0.01$   &$3.25\pm 0.17$&  $3.21\pm0.07$&$2.53\pm0.01$ 
\\ \hline
EQ J$0443-037$  & $6.51$  & $9.12\pm0.12$ & $1.33\pm0.07$ & $3.65\pm 0.24$ & $1.31\pm0.02$   &$3.62\pm0.47$ & $5.47\pm0.35$&$2.56\pm0.02$ 
\\ \hline
EQ J$0916-197$  & $5.64$  & $9.12\pm0.20$ &  $1.33\pm 0.11$ &$3.65\pm0.41$& $1.29\pm0.01$   &$3.60\pm0.78$  &$4.59\pm0.44$&$2.55\pm 0.04$ 
\\ \hline
WD $1659+440$  & $3.05$  & $9.38\pm0.07$ & - &-&   $1.33\pm0.01$   &$2.70\pm0.21$&  $1.96\pm0.07$&$2.50\pm0.01$ 
\\ \hline
EUVE $J1535-77.4$  & $5.80$  & $9.14\pm0.07$ & $1.34\pm0.11$ & $3.58\pm 0.40$ & $1.30\pm0.04$   &$3.52\pm0.76$&  $4.70\pm0.47$& $2.55\pm0.04$
\\ \hline
\enddata
\tablecomments{The mass, radius and central temperature are obtained by a fitting in the effective temperature and gravity for $M_{H}=10^{-4}M_\odot$ and $M_{He}=10^{-2}M_\odot$.}
{{\bf Reference.} \cite{Vennes_1997}.}
\end{deluxetable*}

{ In Table \ref{tabledata1_v3}, we analyze the same white dwarfs shown in Table \ref{tabledata1_v2} but considering $M_{He}=10^{-2}$ and $M_{H}=10^{-4}$. The results in mass and radius are very similar to the ones found in \citep{Vennes_1997}, except for WD $0346-011$. This is due to the effects of general relativity being more notorious at large total masses. For the stars in Table \ref{tabledata1_v3}, we found  $\alpha=2.54\pm0.02$. This very similar to the one reported in \citep{Koester_1976} and used by \cite{Boshkayev_2016}, i.e.,  $\alpha=2.56$. The central temperatures reported in this table are smaller than the ones placed in Table \ref{tabledata1_v2}, due to photons' luminosity that increases with the hydrogen contribution.} { In Tables \ref{tabledata1_v2} and \ref{tabledata1_v3}, we find that all massive observable white dwarfs have masses below $M\leq 1.35M_\odot$. In a cooling process, these stars would not run from stable to unstable regions. In contrast, they are likely to cool in a common white dwarf evolutionary track.}

\begin{figure}[h!] 
\begin{center}
\includegraphics[width=0.6\linewidth]{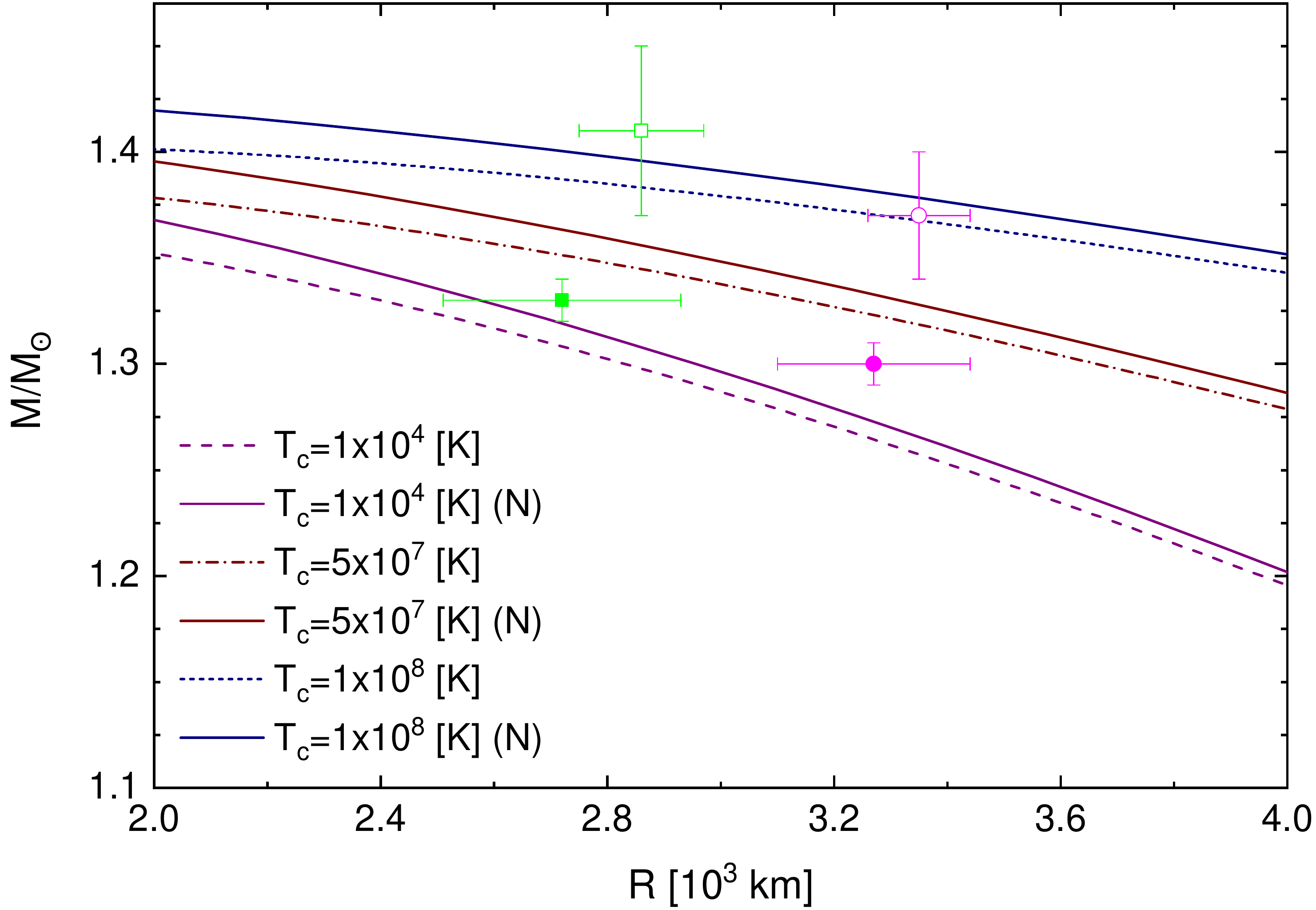}
\caption{The mass against the total radius for some temperatures according to general relativity (dashed lines) and Newtonian formulations (full lines). The green points represent WD $1659+440$ with helium envelope and the pink ones WD $0346-011$ for a helium and hydrogen envelope. The unfilled points are obtained according to \citep{Vennes_1997} and the filled ones with our fitting.}
\label{fig_fit}
\end{center}
\end{figure}

In Figure \ref{fig_fit} we show the mass as a function of radius for some temperatures. The dashed curves are the same ones obtained in Figure \ref{fig1} considering a TOV relativistic equation. The continuous curves are obtained using the Newtonian formulation \citep{Carvalho_2018}. The green points represent the source WD $1659+440$ with a helium envelope and the pink ones WD $0346-011$ for a helium and hydrogen envelope. The unfilled points are obtained according to \citep{Vennes_1997} and the filled ones with our fitting.

From Fig. \ref{fig_fit}, as obtained in \citep{Carvalho_2018} for cold white dwarfs, at the range of maximum white dwarfs masses, for a fixed central temperature $T_c\neq0$, we obtain smaller masses at the GR scope than in the Newtonian formulations. In addition, when the central temperature is increased, the total mass stays closer to white dwarfs WD $1659+440$ and WD $0346-011$ masses. In turn, at a fixed total mass, it is found a lower total radius in the relativistic case than in the classical one.

{ We propose an equation that relates mass and gravity values according to the central temperature. For such equation, we fit the curves in Fig.~\ref{fig1} in Fourier second-order equations to obtain the relation:}
\begin{align}\label{equation_mg}
    M\left(g\right)= \frac{k_0\left[\log\left(\frac{g}{g_\odot}\right)\right]^2+k_1\left[\log\left(\frac{g}{g_\odot}\right)\right]+k_2}{\log\left(\frac{g}{g_\odot}\right)+k_3},
\end{align}
{ being $g_\odot$ the Sun surface gravity, $k_0$, $k_1$, and $k_2$ are parameters in Solar masses $M_{\odot}$, and $k_3$ is a dimensionless constant. This relation is valid for surface gravity values $ \log\left(\frac{g}{g_\odot}\right)\geq 4.4$. The fits implemented have a ${\rm R-square}=1$. As shown in Table \ref{table4}, the parameters $k_0$, $k_1$,  $k_2$, and $k_3$ depend on the central temperature $T_c$. For different $T_c$ than those ones reported in Table \ref{table4}, new numerical values for parameters from Eq. \eqref{equation_mg} can be obtained by interpolating the curves $M(g)$ from Fig.~\ref{fig1}.}

\begin{deluxetable*}{c|c|c|c|c}
\tablenum{4}
\tablecaption{Central temperatures and the parameter values appearing in Eq. \eqref{equation_mg}. \label{table4}}
\tablewidth{0pt}
\tablehead{
\colhead{$T_c$} \vline& \multicolumn{4}{c}{Parameters}\\ \cline{2-5}
\colhead{[K]}\vline & $k_0 \rm~ [M_\odot]$ &$k_1 \rm ~[M_\odot]$& $k_2 \rm ~[10 M_\odot]$ & $k_3$ 
}
\startdata
$1.0\times10^6$  & $-0.355$  & $5.400$  & $-1.442$ & $-2.187$  \\ \hline
$1.0\times10^7$  & $-0.347$  & $5.293$  & $-1.402$ & $-2.162$  \\ \hline
$5.0\times10^7$  & $-0.333$  & $5.076$  & $-1.277$ & $-1.845$ \\ \hline
$1.0\times10^8$  & $-0.324$  & $4.894$  & $-1.137$ & $-1.414$ \\ \hline
\enddata
\end{deluxetable*}

\subsection{Stability of hot white dwarfs}\label{section_stability}

\subsubsection{Stability of hot white dwarfs against small radial perturbations}

\begin{figure}[h]
\center
\includegraphics[width=0.6\linewidth]{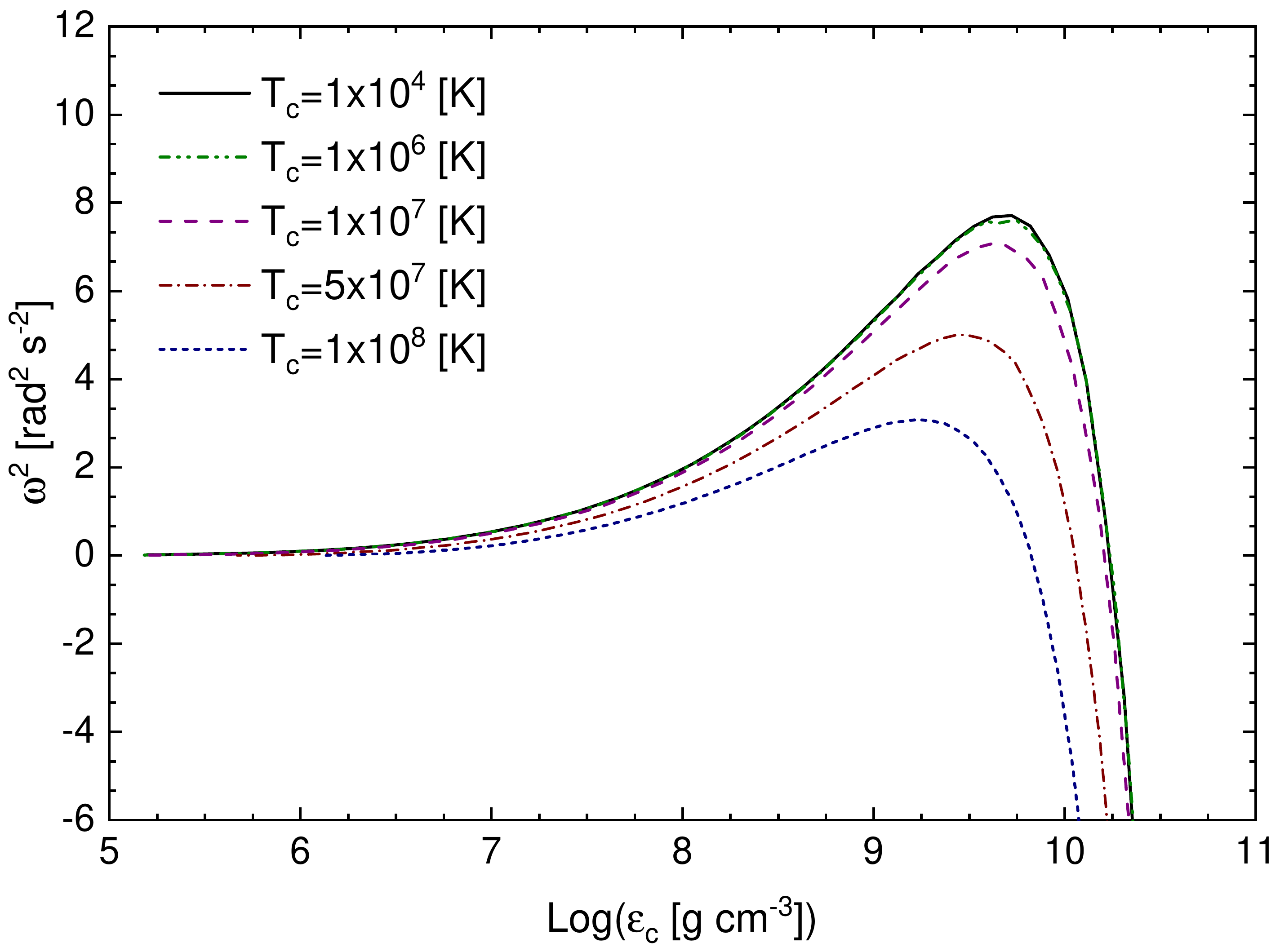}
\includegraphics[width=0.6\linewidth]{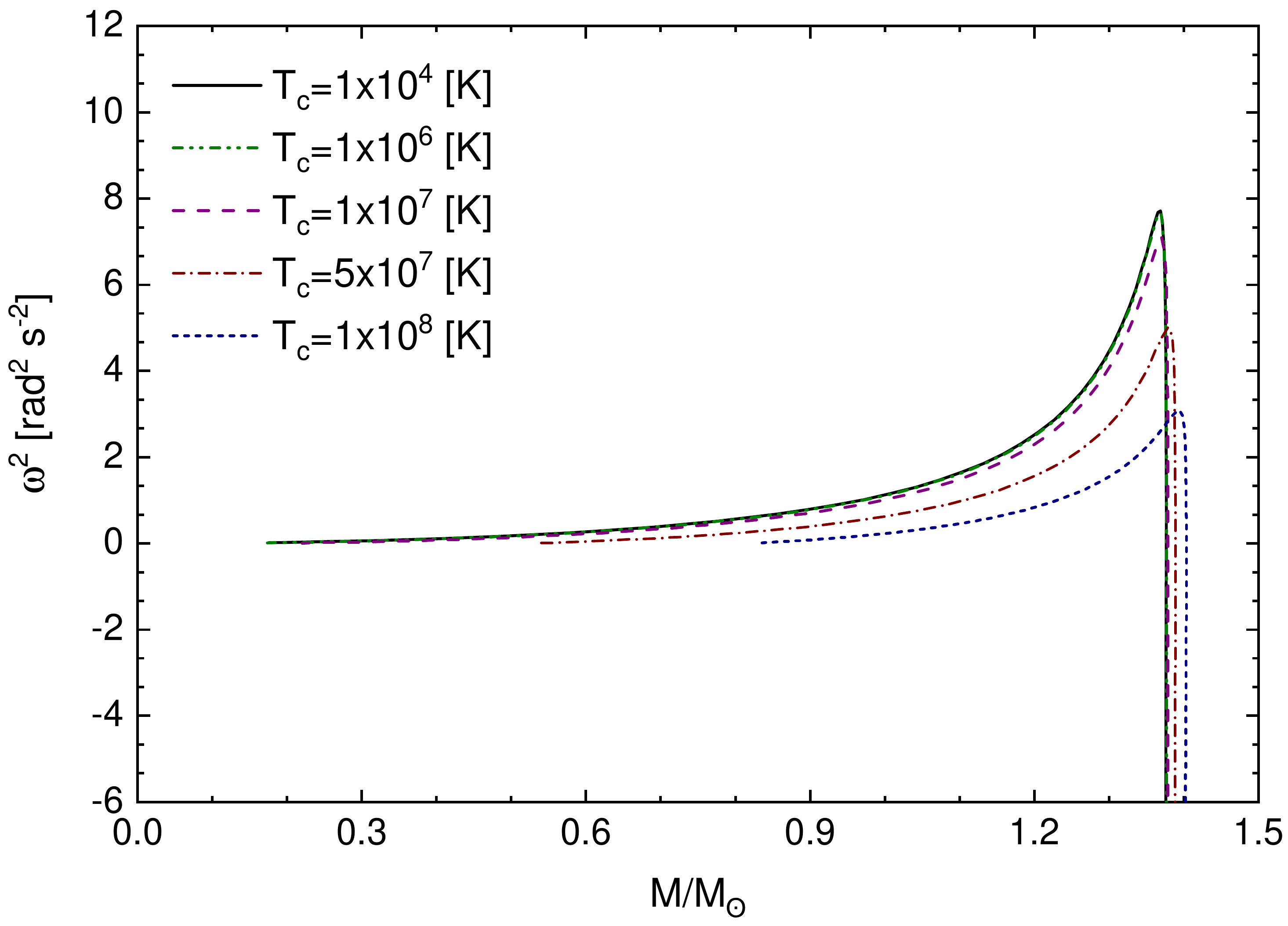}
\caption{Square of eigenfrequency $\omega^2$ as a function of the central energy $\varepsilon_c$ and of the total mass are shown on the panels on top and bottom, respectively, for some central temperatures.}\label{fig6}
\end{figure}

{ The very massive white dwarfs reported in \citep{Vennes_1997} have masses in the range of the instabilities by radial oscillations, inverse $\beta$-decay, and pycnonuclear reactions. This with their central temperature that reaches very high values, near $10^8\rm ~[K]$, made us assume the importance of investigating the stability of hot white dwarfs.}

The behavior of the eigenfrequency squared with the central energy density $\omega^2(\varepsilon_c)$ and with the total mass, $\omega^2(M/M_{\odot})$ are plotted on the top and bottom of Fig.~\ref{fig6}, respectively, for some central temperatures. All curves $\omega^2(\varepsilon_c)$ and $\omega^2(M/M_{\odot})$ show a Gaussian behavior, where the heights of the curves' peaks are respectively  found in the range of central energy densities $1.0\times10^{8}\lesssim\varepsilon_c\lesssim5.0\times10^{10}[\rm g~cm^{-3}]$ and total masses $1.20\lesssim M/M_{\odot}\lesssim1.40$. From the curves $\omega(M/M_{\odot})$, it can be seen that the points of maximum mass are reached at $\omega^2=0$. In addition, it is important to say that the curve $\omega^2(\varepsilon_c)$ derived for $T_c\leq10^6 \rm [K]$, are similar to the one reported by \cite*{Wheeler_1968} and \cite{Chanmugam_1977}. 

The influence of temperature on the radial stability can be also observed in Fig.~\ref{fig6}. In both panels in figure, it can be noted that the increment of $T_c$ decreases the squared eigenfrequency of the fundamental mode, this indicates that hotter white dwarf will have lower stability. { In fact, in our fitting for the stars reported in Tables \ref{tabledata1_v2} and \ref{tabledata1_v3} we also calculated their fundamental eigenfrequency. For white dwarfs composed by $M_{He}=10^{-4}M_\odot$ we found $2.5 \leq{\omega^2}\leq4.4\;\rm [rad^2\;s^{-2}]$ and for ones made of $M_{H}=10^{-4}M_\odot$ and $M_{He}=10^{-2}M_\odot$, we found $2.8\leq{\omega^2}\leq4.6\; \rm [rad^2\;s^{-2}]$. Furthermore, most of the observable white dwarfs reported in \citep{Vennes_1997,Madej_2004,Nalez_2004,Tremblay_2011,Koester_2019}, i.e., white dwarfs with masses between $0.3\leq M/M_\odot\leq1.3$,
have the eigenfrequency of oscillation in the interval $0<\omega^2\leq4.5 \rm \;[rad^2 \;s^{-2}]$.}

\subsubsection{Stability of hot white dwarfs against pycnonuclear reactions and inverse $\beta$-decay}

Recently, Otoniel and collaborators in \citep{Otoniel_2019} discussed that the threshold central energy density, at which pycnonuclear reactions occur, is obtained by taking into account $\tau_{pyc}=10\,[\rm Gyr]$. In our model, at zero temperature, pycnonuclear reactions occur at the threshold density of $9.56\times10^{9}[\rm g~cm^{-3}]$, being a very close value to the one derived in \citep{Otoniel_2019}. In turn, unlike pycnonuclear reactions, the threshold of central density for inverse $\beta$-decay instabilities is $3.52\times10^{10}[\rm g~cm^{-3}]$, close to ones derived in \citep{Rotondo_2011,Otoniel_2019}. 

\begin{figure}[h!]
\begin{center}
\includegraphics[width=0.6\linewidth]{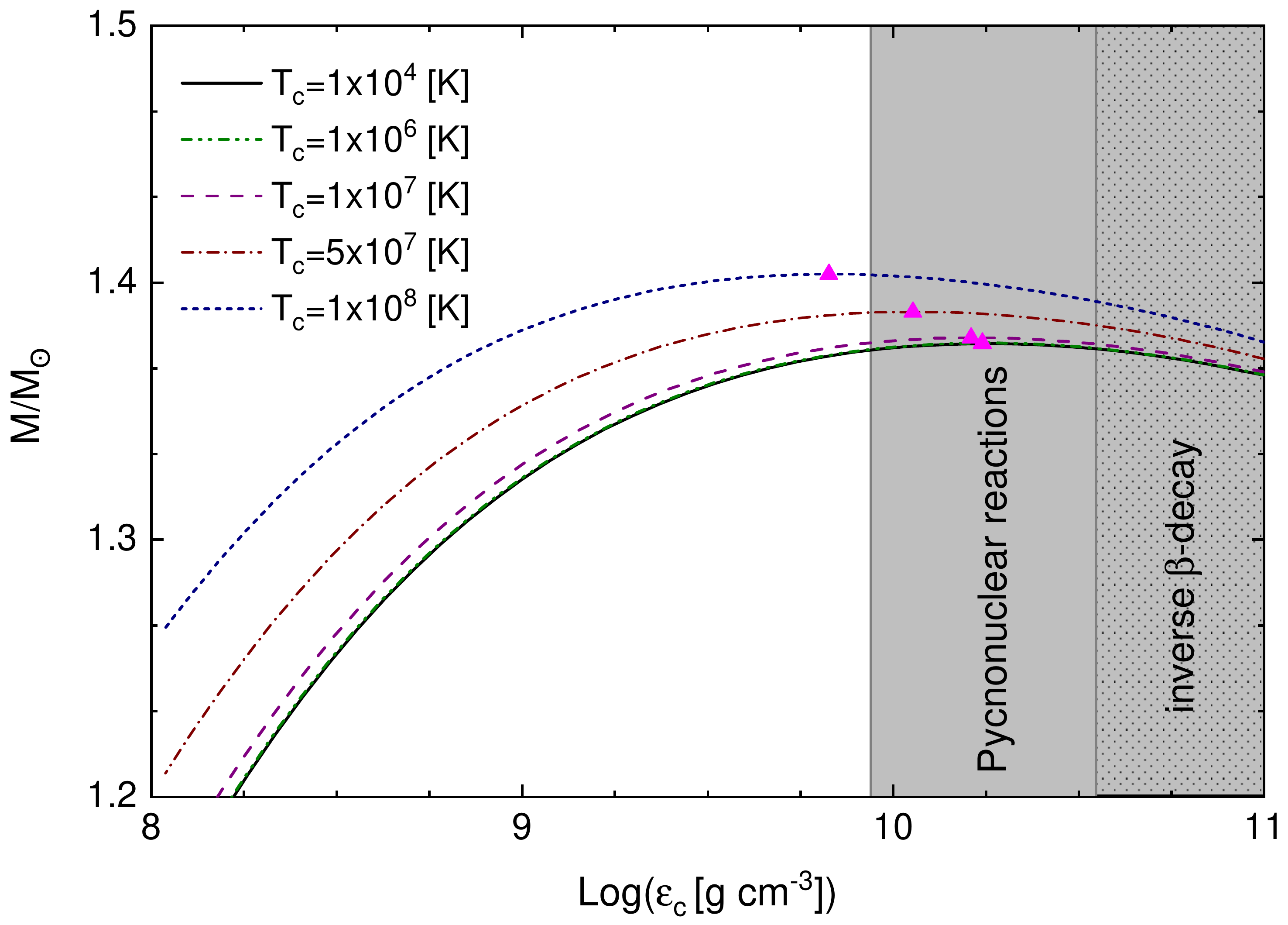}
\caption{Total mass against the central energy density for some central temperatures. The gray shaded regions indicate the places where the instabilities against pycnonuclear reactions and inverse $\beta$-decay take place, and the full triangle on pink mark the onset of the radial instability.}
\label{fig8}
\end{center}
\end{figure}

The behavior of the total mass with the central energy density, at large total masses, it is shown in Fig. \ref{fig8} for some central temperatures. In the figure, we present thresholds where occurs the instabilities against pycnonuclear reactions and inverse $\beta$-decay, in the gray shaded regions, and the place where the radial instability begins, marked by pink full triangles. 

\begin{deluxetable*}{cccc}
\tablenum{5}
\tablecaption{Threshold energy density values for instability against pycnonuclear reactions, inverse $\beta$-decay, and radial oscillations for some central temperature values.\label{table4data}}
\tablewidth{0pt}
\tablehead{
\colhead{$T$} & \colhead{$\varepsilon_{pyc}^{*}$} & \colhead{$\varepsilon_{\beta}^{*}$} & \colhead{$\varepsilon_{\omega}^{*}$}  \\
\colhead{$10^{7}[\rm K]$} & \colhead{$10^{10}[\rm g~ cm^{-3}]$} & \colhead{$10^{10}[\rm g ~cm^{-3}]$} & \colhead{$10^{10}[\rm g ~cm^{-3}]$} 
}
\startdata
$0.1$ & $0.878$ & $ 3.515$ & $1.744$ 
\\ \hline
$1.0$ & $0.878$ & $ 3.515$ & $1.622$ 
\\ \hline
$5.0$ & $0.876$ & $ 3.515$ & $1.118$ 
\\ \hline
$10$ & $0.874$ & $ 3.515$ & $0.675$ 
\enddata
\end{deluxetable*}


Table \ref{table4data} shows threshold energy density values for instability against pycnonuclear reactions, inverse $\beta$-decay, and radial oscillations for four central temperature values. From the results, we can note that the increase of temperature affects more notorious stability against small radial perturbations. Moreover, for central temperature larger than $1\times10^8\,[\rm K]$, the radial stability is attained before the ones produced by pycnonuclear reactions and inverse $\beta$-decay.

\section{Conclusions}\label{conclusions}

In this article, the static equilibrium configuration and stability against small radial perturbation, pycnonuclear reaction, and inverse $\beta$-decay in white dwarfs with a finite temperature were studied. Following in \citep{Tolman_1939,Timmes_1999,Rotondo_2011}, with respect to the matter within white dwarfs,
we take into account that it is made of nucleons and electrons confined in a Wigner-Seitz cell surrounded by free photons. Moreover, with the purpose to obtain a null radiation pressure at the star's surface, a temperature distribution was considered in the non-degenerate envelope.  We assume both that temperature depends on the mass density and the existence of an isothermal degenerate core. The static configurations under analysis have spherical symmetry and are connected smoothly with the Schwarzschild exterior spacetime. The hydrostatic configuration and stability were investigated for different central energy densities $\varepsilon_c$ and central temperatures $T_c$. 

We obtained that the necessary central temperature to influence the static structure and radial stability of very massive white dwarfs is approximately $T_c=10^7\,[\rm K]$. For a fixed central energy density, we found that white dwarfs' radius and total mass grow with the increase of the central temperature (for $T_c>10^{7}\,[\rm K]$). The temperature effects in the static equilibrium configurations are in concordance with ones obtained in the study of white dwarfs with a finite temperature reported in \citep{deCarvalho_2014} for low mass white dwarfs but are different from the ones obtained in \citep{Boshkayev_2016} for the very massive ones. { Analyzing the mass and radius according to the central temperature reported in this work, we can understand that hot ($T_c>10^7\rm[K]$) and massive white dwarfs ($M>1.37M_\odot$) could collapse. In this sense, further studies of cooling of massive white dwarfs considering general relativity and instabilities due to radial perturbations, pycnonuclear reactions, and inverse $\beta$-decay  are needed to conclude if they are likely to cool or collapse.
}

{ For some massive white dwarfs, we derive their mass and radius by fitting both gravity and effective temperature reported by the observation. This is done by assuming white dwarfs are composed by $M_{He}=10^{-4}M_{\odot}$, and other ones made of $M_{H}=10^{-4}M_{\odot}$ and $M_{He}=10^{-2}M_{\odot}$. The results show that some white dwarfs could have central temperatures above $5\times10^7 \rm [K]$. Besides, from our results, we note that masses and radius are in the same range of those white dwarfs reported in \citep{Vennes_1997}, with exception of EUVE J$0003+43.6$ and WD $1659+440$. For these two stars, we found masses below those ones reported in \citep{Vennes_1997}. This is associated with the relativistic effects on these two massive stars. Furthermore, considering general relativity effects, we derive an equation that facilitates finding mass values from surface gravity and effective temperature values for observable massive white dwarfs with surface gravity  $ \log\left(\frac{g}{g_\odot}\right)\geq 4.4$. }

{ We made a novel study concerning the stability of white dwarfs with a finite temperature.} For a central energy density interval, we found that the radial stability of the white dwarfs diminishes with the increment of the central temperature. Moreover, we also derived that the maximum mass and the zero eigenfrequencies of oscillation are attained at the same central energy density. This point out that in a system of equilibrium stars configurations { at finite temperature }, the regions formed by stable and unstable white dwarfs can be recognized by the inequalities $dM/d\varepsilon_c>0$ and $dM/d\varepsilon_c<0$, respectively.

On the other hand, unlike the threshold energy density of instability for inverse $\beta$-decay, which does not change with temperature, the energy density threshold of instability for pycnonuclear fusion reaction is reduced when the temperature in the stellar interior is present. Besides, for central temperature larger than $1\times10^8\,[\rm K]$, we determined that instability due to small radial perturbations occur before those produced by pycnonuclear reactions and inverse $\beta$-decay.



\begin{acknowledgments}

\noindent We would like to thank Funda\c{c}\~ao de Amparo \`a Pesquisa do Estado de S\~ao Paulo (FAPESP), Grant No. $2013/26258-4$. SPN thanks Conselho Nacional de Desenvolvimento Cient\'ifico e Tecnol\'ogico (CNPq), Grant No. 	$140863/2017-6$ for the financial support. JDVA would like to thank the Universidad Privada del Norte and Universidad Nacional Mayor de San Marcos for funding - RR Nº$\,005753$-$2021$-R$/$UNMSM under the project number B$21131781$. MM is grateful to CAPES and CNPq financial support.

\end{acknowledgments}

\bibliography{bibliography}{}
\bibliographystyle{aasjournal}



\end{document}